\documentclass[english,showpacs,pra]{revtex4-1}
\usepackage[T1]{fontenc}
\usepackage[latin9]{inputenc}
\usepackage{amsmath}
\usepackage{amssymb}
\usepackage{babel}

\begin{document}

\title{Nonexistence of a universal quantum machine to examine the precision
of unknown quantum states}

\author{Shengshi Pang, Shengjun Wu and Zeng-Bing Chen}

\affiliation{Hefei National Laboratory for Physical Sciences at Microscale and
Department of Modern Physics, University of Science and Technology
of China, Hefei, Anhui 230026, China}
\begin{abstract}
In this work, we reveal a new type of impossibility discovered in
our recent research which forbids comparing the closeness of multiple
unknown quantum states with any non-trivial threshold in a perfect
or an unambiguous way. This impossibility is distinct from the existing
impossibilities in that it is a ``collective'' impossibility on
multiple quantum states while most other ``no-go'' theorems concern
with only one single state each time, i.e., it is an impossibility
on a non-local quantum operation. This novel impossibility may provide
a new insight into the nature of quantum mechanics and it implies
more limitations on quantum information tasks than the existing ``no-go''
theorems.
\end{abstract}


\maketitle
Quantum mechanics has brought many surprises to people with its fantastic
features and broad applications for a long time since it was born.
In recent decades, it has been applied to the information field and
greatly furthered this field by introducing the concepts like nonlocality
and entanglement \cite{entanglement} which lead to the emergence
of some novel quantum protocols and algorithms such as quantum teleportation
\cite{teleportation}, quantum dense coding \cite{dense coding},
Shor's factoring algorithm \cite{shor}, etc, demonstrating higher
performance in communication and computation than classical protocols.

On the other hand, quantum mechanics has also laid its distinctive
limitations on quantum information tasks due to its linearity and
superposition principle, such as the well-known quantum no cloning
theorem \cite{nocloning}, the quantum no deletion theorem \cite{nodeleting}
and so on. More recently, some new no-go theorems are found, including
the no-broadcasting theorem \cite{no broadcasting,no broadcasting2},
the impossibility of unconditionally secure quantum bit commitment
\cite{bit comit 4,bit comit2,bit comit3,bit commit}, the impossibility
of sequential implementation of global unitary operations with an
itinerant ancillary system \cite{Sequential,sequential2}, the nonexistence
of deterministic purification for two copies of a noisy entangled
state \cite{purification}, and so on. These impossibilities have
provided deep insights into the quantum world and have stimulated
people to explore more intrinsic nature of the quantum mechanics.

In this article, we are going to study a problem concerning the ``closeness''
of multiple unknown quantum states and reveal a interesting collective
type of impossibility in quantum mechanics.

Consider that when a classical machine produces many copies of a product,
people may control the quality of the products by using all kinds
of apparatus to find whether the copies of the same product are exactly
identical or the difference between them is within a tolerable level.
In the quantum world, one may also try to find whether or not several
quantum states are the same or the difference between them is within
a threshold by quantum measurement to control the quality of the quantum
states or test the stability of quantum machines. However, our research
in this paper implies that such a task is forbidden in the quantum
world. In this work, we shall show that with a sound definition of
\emph{closeness }between multiple quantum states, one cannot determine
whether the closeness of the unknown states is within or beyond a
given tolerable threshold $A$.

The problem is defined as follows: suppose there are $n$ unknown
states $|\psi_{1}\rangle,\cdots,|\psi_{n}\rangle$ arbitrarily chosen
from the Hilbert space $\mathcal{H}$ of finite dimension $d$ (any
state in $\mathcal{H}$ can be chosen repeatedly for arbitrary times),
and we use the average fidelity between them,
\begin{equation}
C=\frac{2}{n(n-1)}\sum_{i<j}\left|\langle\psi_{i}|\psi_{j}\rangle\right|^{2},\label{eq:0}
\end{equation}
to measure how close the $n$ states are, then we ask ``is it possible
to determine whether or not the average fidelity of the $n$ unknown
states is above a given threshold by quantum measurements?'' If we
denote the threshold as $A$, then this task is to test the inequality
\begin{equation}
C\geq A,\label{eq:1}
\end{equation}
by quantum measurements. We shall call $C$ the \emph{closeness} of
$|\psi_{1}\rangle,\cdots,|\psi_{n}\rangle$ in this paper.

Here, we need to note that there is some natural limit on $C$. When
$n\leq d$, the minimum of $C$ is $0$ obviously, so $A$ can take
any value between $0$ and $1$. But when $n>d$, there are no $n$
orthogonal states in the Hilbert space $\mathcal{H}$, so the closeness
$C$ has a non-zero minimum value in this case, and we denote it as
$C_{\mathrm{min}}$. In the appendix, we derive the minimum value
of $C$:
\begin{equation}
C_{\mathrm{min}}=\frac{n-d}{d(n-1)},\; n>d.\label{eq:17}
\end{equation}
Let $C_{\mathrm{min}}=0$ when $n\leq d$. We assume $C_{\mathrm{min}}\leq A\leq1$
throughout this paper, whether $n\leq d$ or not, otherwise no measurement
is necessary, and the problem would be trivial.

In our research, we find that the above task is definitely impossible
to accomplish by a perfect or an unambiguous measurement, whenever
$C_{\mathrm{min}}<A<1$. ``Perfect measurement'' means that the
measurement will deterministically produce a conclusive and correct
result, and ``unambiguous measurement'' means that the measurement
may produce an inclusive result with non-zero probability, but if
a result is conclusive, it must be error-free. For a review of unambiguous
quantum measurement, we refer the readers to \cite{referee}. We shall
focus on the unambiguous measurement in this article, since the perfect
measurement is a special case of the unambiguous measurement.

At first glance, the impossibility of comparing how close several
unknown quantum states with a given threshold does not seem strange,
since it is widely known that an unknown quantum state cannot be determined
generally due to the no cloning theorem, then the average fidelity
between the unknown states cannot be determined, either. However,
it should be clear that what is actually concerned in our problem
is the relation between the unknown states, but not what each state
is, and it is not necessary to determine each state first so as to
acquire the relation between them. In fact, collective measurement,
one of the characteristic operations in quantum mechanics which cannot
be decomposed to local measurements on single states generally, can
be performed on the $n$ unknown states simultaneously, implying more
success possibility of discerning the relation between them than that
of determining each unknown quantum state.

Our research result is two-folded in detail: i) if $C_{\mathrm{min}}<A\le1$,
no measurement can unambiguously indicate the case that the average
fidelity of the unknown quantum states is above the threshold; ii)
if $C_{\mathrm{min}}\leq A<1$, no measurement can unambiguously indicate
the case that the average fidelity of the unknown quantum states is
below the threshold.

We use the \emph{positive operator-valued measure} (POVM) \cite{nielsen}
to study the problem in our research. POVM provides a convenient way
to describe a general physical process (no matter the process is local
or non-local) if only the statistical properties of the process are
concerned in the problem. A POVM consists of a set of POVM elements,
each of which is a positive operator and corresponds to a possible
outcome of the physical process, and the POVM elements sum up to the
identity operator on the Hilbert space. In our research, there are
three possible outcomes after measuring the closeness of the given
states: i) the average fidelity between the states is above the given
threshold, ii) the average fidelity between the states is below the
given threshold, and iii) the result is inconclusive. We use $R_{1}$,
$R_{2}$, $R_{?}$ to denote the three possible result respectively
and $M_{1}$, $M_{2}$ and $M_{?}$ to denote the corresponding POVM
elements. Each $M_{i}$ acts on the composite Hilbert space $\mathcal{H}^{\otimes n}$
of the $n$ quantum states as simultaneous measurements are allowed
on the whole $n$ states in this problem. Quantum mechanics tells
that the probability the result $M_{i}$ occurs is
\begin{equation}
\mathrm{Prob}(R_{i})=\langle\psi_{1}|\otimes\cdots\otimes\langle\psi_{n}|M_{i}|\psi_{1}\rangle\otimes\cdots\otimes|\psi_{n}\rangle,\label{eq:2}
\end{equation}
and the unambiguity of the measurement requires that $\mathrm{Prob}(R_{1})=0$
if the average fidelity of the $n$ states is below the threshold
$A$ and $\mathrm{Prob}(R_{2})=0$ if the average fidelity of the
$n$ states is above the threshold.

Now we present the proof of our result. Let us divide all product
states in the composite Hilbert space $\mathcal{H}^{\otimes n}$ into
two sets: one set contains all product states whose $n$ factor states
satisfy \eqref{eq:1}, and the other set contains the remaining product
states. We denote the first set by $S_{1}$ and the second set by
$S_{2}$. Then, the task of examining the closeness of $n$ arbitrary
quantum states with a threshold $A$ is equivalent to distinguishing
between the sets $S_{1}$ and $S_{2}$.

We first prove the first part of our result: that the average fidelity
of the unknown quantum states is above the threshold cannot be detected
unambiguously if $C_{\mathrm{min}}<A\le1$. The core of the proof
is to show $S_{2}$ can span the whole composite Hilbert space $\mathcal{H}^{\otimes n}$.
We prove this by constructing a spanning set of $\mathcal{H}^{\otimes n}$
from the set $S_{2}$.

Let us arbitrarily select $n$ quantum states $|\psi_{1}\rangle,\cdots,|\psi_{n}\rangle$,
of which the average fidelity $C$ \eqref{eq:0} is below the threshold
from the Hilbert space $\mathcal{H}$, then $|\psi_{1}\rangle\otimes\cdots\otimes|\psi_{n}\rangle\in S_{2}$.
Suppose $|\phi_{i,j}\rangle$, $j=1,\cdots,d-1$, are $d-1$ orthonormal
basis states of the orthogonal complement to $|\psi_{i}\rangle$ in
$\mathcal{H}$. Let
\begin{equation}
|\psi_{i,j}\rangle=\frac{1}{\sqrt{1+|\epsilon|^{2}}}\left(|\psi_{i}\rangle+\epsilon|\phi_{i,j}\rangle\right),\: j=1,\cdots,d-1,\label{eq:6}
\end{equation}
and $|\psi_{i,d}\rangle=|\psi_{i}\rangle$, for each $i=1,\cdots,n$,
it is evident that the $d$ states $|\psi_{i,j}\rangle$, $j=1,\cdots,d$
are linearly independent. When $\epsilon$ is sufficiently small,
for arbitrary $n$ indexes $j_{1},\cdots,j_{n}=1,\cdots,d$ the average
fidelity $C^{\prime}$ between the states $|\psi_{1,j_{1}}\rangle,\cdots,|\psi_{n,j_{n}}\rangle$
becomes
\begin{equation}
C^{\prime}=\frac{2}{n(n-1)(1+|\epsilon|^{2})}\sum_{k<l}|\langle\psi_{k}|\psi_{l}\rangle+\epsilon(\langle\psi_{k}|\phi_{l,j_{l}}\rangle+\langle\phi_{k,j_{k}}|\psi_{l}\rangle)+\epsilon^{2}\langle\phi_{k,j_{k}}|\phi_{l,j_{l}}\rangle|^{2}.\label{eq:11}
\end{equation}
It can be seen that when $\epsilon$ is sufficiently small, $C^{\prime}$
can be still below the threshold: note that $C^{\prime}|_{\epsilon=0}=C$,
so when $\epsilon$ is sufficiently small, $|C^{\prime}-C|\approx|\frac{dC^{\prime}}{d\epsilon}|_{\epsilon=0}||\epsilon|<A-C$
if $|\epsilon|<|A-C||\frac{dC^{\prime}}{d\epsilon}|_{\epsilon=0}|^{-1}$,
then $C^{\prime}\leq C+|C^{\prime}-C|<C+A-C=A$, thus $|\psi_{1,j_{1}}\rangle\otimes\cdots\otimes|\psi_{n,j_{n}}\rangle\in S_{2}$.
Now we show that such $d^{n}$ states $|\psi_{1,j_{1}}\rangle\otimes\cdots\otimes|\psi_{n,j_{n}}\rangle$
forms a spanning set of $\mathcal{H}^{\otimes n}$ due to the linear
independence of $|\psi_{i,j}\rangle$, $j=1,\cdots d$ for each $i$.
Let $|\Psi_{1}\rangle\otimes\cdots\otimes|\Psi_{n}\rangle$ be an
arbitrary product state in $\mathcal{H}^{\otimes n}$. Since the $d$
states $|\psi_{i,j}\rangle$, $j=1,\cdots,d$ are linearly independent
for each $i=1,\cdots,n$ and the dimension of $\mathcal{H}$ is $d$,
the state $|\Psi_{1}\rangle\otimes\cdots\otimes|\Psi_{n}\rangle$
can be expanded as
\begin{equation}
|\Psi_{1}\rangle\otimes\cdots\otimes|\Psi_{n}\rangle=\sum_{j_{1},\cdots,j_{n}=1}^{d}\alpha_{j_{1}}\cdots\alpha_{j_{n}}|\psi_{1,j_{1}}\rangle,\cdots,|\psi_{n,j_{n}}\rangle.\label{eq:5}
\end{equation}
Considering that any state in $\mathcal{H}^{\otimes n}$ can be expanded
by product states, the whole composite Hilbert space $\mathcal{H}^{\otimes n}$
can be spanned by the states $|\psi_{1,j_{1}}\rangle\otimes\cdots\otimes|\psi_{n,j_{n}}\rangle,\: j_{1},\cdots,j_{n}=1,\cdots,d$
then, so the $d^{n}$ states $|\psi_{1,j_{1}}\rangle\otimes\cdots\otimes|\psi_{n,j_{n}}\rangle$,
$j_{1},\cdots,j_{n}=1,\cdots,d$ forms a basis of $\mathcal{H}^{\otimes n}$.

In the following, we show that in any case the probability of producing
the outcome $R_{1}$ by an unambiguous measurement must be zero. Since
the two operators $M_{1}$ and $M_{2}$ are positive, they can be
decomposed as
\begin{equation}
M_{i}=K_{i}^{\dagger}K_{i},\quad i=1,2.\label{eq:3}
\end{equation}
The unambiguity of the measurement requires that the outcome $R_{1}$
should not occur when the average fidelity of the $n$ unknown states
is below the threshold, i.e.
\begin{equation}
\langle\psi_{1,j_{1}}|\otimes\cdots\otimes\langle\psi_{n,j_{n}}|K_{1}^{\dagger}K_{1}|\psi_{1,j_{1}}\rangle\otimes\cdots\otimes|\psi_{n,j_{n}}\rangle=\left\Vert K_{i}|\psi_{1,j_{1}}\rangle\otimes\cdots\otimes|\psi_{n,j_{n}}\rangle\right\Vert ^{2}=0,\,\forall|\psi_{1,j_{1}}\rangle\otimes\cdots\otimes|\psi_{n,j_{n}}\rangle\in S_{2},\label{eq:4}
\end{equation}
then we have
\begin{equation}
K_{1}|\psi_{1,j_{1}}\rangle\otimes\cdots\otimes|\psi_{n,j_{n}}\rangle=0,\label{eq:7}
\end{equation}
thus
\begin{equation}
M_{1}|\psi_{1,j_{1}}\rangle\otimes\cdots\otimes|\psi_{n,j_{n}}\rangle=K_{1}^{\dagger}K_{1}|\psi_{1,j_{1}}\rangle\otimes\cdots\otimes|\psi_{n,j_{n}}\rangle=0.\label{eq:8}
\end{equation}
For arbitrary $n$ quantum states $|\Psi_{1}\rangle,\cdots,|\Psi_{n}\rangle$,
 \eqref{eq:5} implies that
\begin{equation}
M_{1}|\Psi_{1}\rangle\otimes\cdots\otimes|\Psi_{n}\rangle=0,\label{eq:9}
\end{equation}
resulting in
\begin{equation}
M_{1}=0.\label{eq:10}
\end{equation}
Therefore, it can be inferred that the outcome $R_{1}$, which indicates
that the average fidelity of the unknown states is above the threshold,
will never be produced in any case if the measurement is unambiguous.
Note that the case $A=0$ is excluded in the first part of our result,
because the fidelity between any two quantum states is always non-negative,
a trivial case.

The second part of our result, i.e., no quantum measurement can give
an unambiguous result when the average fidelity of the $n$ quantum
states is below the threshold $A$ if $C_{\mathrm{min}}\leq A<1$,
can be proved in a similar way as above. We skip the details of the
proof here.

Putting the two parts of our result together, we can conclude that
neither $M_{1}$ nor $M_{2}$ exists in any unambiguous quantum measurement
to compare the average fidelity of arbitrary $n$ unknown quantum
states with a threshold $A$ if $C_{\mathrm{min}}<A<1$, and hence
perfect or unambiguous examining the closeness of multiple unknown
quantum states with a threshold $A$ is definitely prohibited when
$C_{\mathrm{min}}<A<1$.

It is worth mentioning that the case $A=1$ is excluded here, because
when $A=1$ our problem is equivalent to determining whether $n$
unknown quantum states are exactly identical. This is reduced to the
problem of quantum state comparison \cite{generalization not universal,ensembles 08,jmo,orginal pla,unambiguous jpa,jinian}
and it can be shown that $M_{2}$ exists (but $M_{1}$ still vanishes)
in this situation, and $M_{2}$ can be chosen as the projector onto
the orthogonal complement to the totally symmetric subspace of $\mathcal{H}^{\otimes n}$
\cite{unambiguous jpa}. In addition, the case $n\leq d$ and $A=0$
is studied as the problem of determining the orthogonality of multiple
quantum states in \cite{orthogonality}.

Furthermore, we want to point out that when $A=1$ although $M_{2}$
exists for unknown pure quantum states, it still vanishes for unknown
mixed quantum states. This is referred to \cite{jinian}.

It is known that there have been some ``no-go'' theorems like the
famous quantum no cloning theorem and quantum no deletion theorem,
and they reveal the limitations in quantum information science due
to quantum principles. Compared with those known impossibilities,
the impossibility of comparing the average fidelity of multiple unknown
quantum states with a threshold in this paper has some interesting
features in the following aspect: most existing ``no-go'' theorems
concern with a single quantum system each time and they forbid local
quantum operations; however, in our research the comparison operation
involves multiple quantum systems simultaneously and the forbidden
quantum measurement is non-local indeed, so the impossibility introduced
in this article is a ``collective'' impossibility.

In summary, we have studied the problem of examining the closeness
of $n$ arbitrary quantum states with a threshold $A$. We have shown
that such a task can never succeed by a perfect or unambiguous quantum
measurement if $C_{\mathrm{min}}<A<1$. This is a new kind of impossibility
other than the existing impossibilities, which may pose new challenges
in practical situations. For example, it implies that it would be
impossible to examine the stability of a deterministic quantum machine
by feeding identical quantum states into the machine and comparing
how close its outputs are with a threshold. We hope that our research
can shed light on further understanding the limitations on quantum
information tasks by quantum principles and provide a deeper insight
into the quantum world.

\section*{Acknowledgment}

This work is supported by the NSFC (Grant No. 11075148 and No. 61125502),
the National Fundamental Research Program (Grant No. 2011CB921300),
the Fundamental Research Funds for the Central Universities, and the
CAS.

\section*{Appendix}

The aim of this appendix is to obtain the minimum of the closeness
when $n>d$.

Note that
\begin{equation}
C=\frac{2}{n(n-1)}\sum_{i<j}|\langle\psi_{i}|\psi_{j}\rangle|^{2}=\frac{1}{n(n-1)}(\mathrm{Tr}G^{2}-n),\label{eq:21}
\end{equation}
where $G$ is the Gram matrix of $|\psi_{1}\rangle,\cdots,|\psi_{n}\rangle$,
i.e. $G_{i,j}=\langle\psi_{i}|\psi_{j}\rangle$.

Suppose the eigenvalues of $G$ are $\lambda_{1}\geq\lambda_{2}\geq\cdots\geq\lambda_{n}$.
As $G$ is positive semi-definite and $\mathrm{rank}G\leq d$, $\lambda_{d+1}=\cdots=\lambda_{n}=0$.
Therefore, we only need to consider the first $d$ eigenvalues of
$G$. As $G_{ii}=1$, $\mathrm{Tr}(G)=\lambda_{1}+\cdots+\lambda_{d}=n$.
According to the inequality
\begin{equation}
\frac{1}{m}\sum_{i=1}^{m}x_{i}\leq\sqrt{\frac{1}{m}\sum_{i=1}^{m}x_{i}^{2}},\; x_{i}\geq0,\, i=1,\cdots,m,\label{eq:19}
\end{equation}
where ``$=$'' holds if $x_{1}=\cdots=x_{m}$, it can be inferred
that
\begin{equation}
\mathrm{Tr}G^{2}=\lambda_{1}^{2}+\cdots+\lambda_{d}^{2}\geq\frac{n^{2}}{d},\label{eq:23}
\end{equation}
so
\begin{equation}
C\geq\frac{n-d}{d(n-1)}.\label{eq:26}
\end{equation}

The minimum value of $\mathrm{Tr}G^{2}$ can be achieved when $\lambda_{1}=\cdots=\lambda_{d}=\frac{n}{d}$,
and the corresponding $|\psi_{1}\rangle,\cdots,|\psi_{n}\rangle$
can be constructed as follows. Suppose the eigenstates of $G$ are
$|0\rangle,\cdots,|n-1\rangle$, then $G=\frac{n}{d}(|0\rangle\langle0|+\cdots|d-1\rangle\langle d-1|)$
when $G$ reaches its minimum. To make all the diagonal elements of
$G$ equal to one, we can change the basis to a mutually unbiased
basis of $\{|0\rangle,\cdots,|n-1\rangle\}$ \cite{mub1,mub2}, e.g.
\begin{equation}
|e_{j}\rangle=\frac{1}{\sqrt{n}}\sum_{k=0}^{n-1}\omega^{jk}|k\rangle,\, j=0,\cdots,n-1,\,\omega=\exp(\mathrm{i}\frac{2\pi}{n}).\label{eq:24}
\end{equation}
Under this basis, $G_{ii}=\langle e_{i}|G|e_{i}\rangle=1$, so a feasible
choice for $|\psi_{1}\rangle,\cdots,|\psi_{n}\rangle$ with which
$\mathrm{Tr}G^{2}$ reaches the minimum is as follows:
\begin{equation}
|\psi_{j}\rangle=\frac{1}{\sqrt{d}}\sum_{k=0}^{d-1}\omega^{jk}|k\rangle,\, j=0,\cdots,n-1.\label{eq:25}
\end{equation}
Note that in \eqref{eq:25}, $\omega$ is still $\exp(\mathrm{i}\frac{2\pi}{n})$,
but not $\exp(\mathrm{i}\frac{2\pi}{d})$.

Therefore, the minimum value of $\mathrm{Tr}G^{2}$ is $\frac{n^{2}}{d}$,
and according to \eqref{eq:21} and \eqref{eq:26}, the minimum value
of $C$, $C_{\mathrm{min}}$, is
\begin{equation}
\frac{n-d}{d(n-1)}.\label{eq:22}
\end{equation}

\end{document}